# Monitoring Spent Nuclear Fuel in a Dry Cask Using Momentum Integrated Muon Scattering Tomography

Junghyun Bae,* Stylianos Chatzidakis *

*School of Nuclear Engineering, Purdue University, West Lafayette, IN 47906, Bae43@purdue.edu, Schatzid@purdue.edu

[leave space for DOI, which will be inserted by ANS]

## INTRODUCTION

Nuclear materials accountability and non-proliferation are among the critical tasks to be addressed for the advancement of nuclear energy in the United States. Monitoring spent nuclear fuel (SNF) is important to continue reliable stewardship of SNF storage. Cosmic ray muons have been acknowledged a promising radiographic tool for monitoring SNF due to their highly penetrative nature and high energy ($10^{-1} – 10^4$ GeV) [1]. For example, a muon with energy 3 GeV can travel approximately 10 m in water ($\rho = 1.0$ g/cm$^3$) and ~0.5 m in uranium ($\rho = 19.1$ g/cm$^3$). On the other hand, the mean free path for high-energy photons is much shorter, approximately 0.25 and 0.01 m for water and uranium, respectively [2], [3]. As a result, cosmic ray muons are more suitable and have been used for imaging large and dense objects, e.g., SNF dry casks [4], [5], Fukushima Daiichi Unit-1 [6], [7], and the Great pyramid of Giza [8], [9].

Despite their potential in various applications, the wide application of cosmic ray muons is limited by the naturally low intensity at sea level. To efficiently utilize cosmic ray muons in engineering applications, two important quantities—trajectory and momentum—must be measured. Although various studies demonstrate that there is significant potential for measuring momentum in muon applications [10], [11], it is still difficult to measure both muon scattering angle and momentum in the field. To fill this critical gap, a muon spectrometer using multi-layer pressurized gas Cherenkov radiators was proposed [12]–[14]. However, existing muon tomographic algorithms were developed assuming monoenergetic muon scattering and are not optimized for a measured poly-energetic momentum spectrum [15].

In this work, we develop and evaluate a momentum integrated muon scattering tomography (MMST) algorithm. We evaluate the algorithm on its capability to identify a missing fuel assembly (FA) from a SNF dry cask. To generate cosmic ray muons and simulate their interactions with SNF dry cask, the Monte-Carlo particle transport simulation code, GEANT4, is used [16], [17]. Our results demonstrate that image resolution using MMST is significantly improved when measuring muon momentum with a resolution of 0.1 GeV/c and it can reduce monitoring time by a factor of 10 when compared to that of a conventional muon imaging technique in terms of systematically finding a missing FA. We anticipate our proposed MMST will enable us to simultaneously measure muon scattering angles and momenta in the field and advance the utilizability of cosmic ray muons in many engineering applications: nuclear material control and accountability, waste management, and non-proliferation.

## MUON SCATTERING TOMOGRAPHY

### Multiple Coulomb Scattering

Because muon is a charged lepton similar to electron and tau, it undergoes consecutive Coulombic collisions with nuclei and electrons. This process is known as multiple Coulomb scattering (MCS) [18]. The sum of muon deflection angles due to MCS can be approximated using Gaussian distribution [19]. In this case, the expected muon scattering angle distribution is given by:

$$\frac{dN}{d\theta} = \frac{1}{\sqrt{2\pi}\sigma_\theta} e^{-\theta^2/2\sigma_\theta^2}, \quad (1)$$

where $\theta$ is the muon scattering angle and $\sigma_\theta$ is the standard deviation. $\sigma_\theta$ is a function of muon momentum, length, and density of scattering medium and it is given by [20]:

$$\sigma_\theta = \frac{13.6 \text{ MeV}}{\beta c p} \sqrt{\frac{X}{X_0}} \left[1 + 0.038 \ln\left(\frac{X}{X_0}\right)\right], \quad (2)$$

where $X$ is the length of scattering medium, $X_0$ is the radiation length, $p$ is the muon momentum, and $\beta c$ is the muon velocity in terms of the speed of light, $c$. Radiation length, $X_0$, is a nuclear property and can be found in the Particle Data Group library [21], [22]. When a muon initially travels along the z-axis, it can be deflected in both x- and y-axes in accordance with MCS Gaussian approximation as schematically shown in Fig. 1. It is possible to estimate target objects by analyzing the muon scattering angle distribution and Gaussian MCS approximation. Feasibility of material estimation is improved by collecting a statistically large number of muon samples because MCS approximation is developed based on a statistical inference technique.

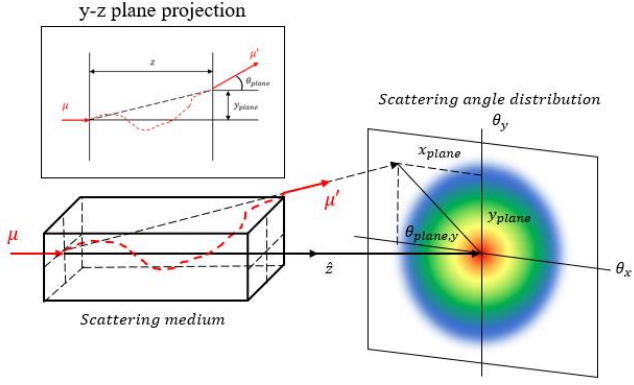

Fig. 1. Multiple Coulomb Scattering (MCS) of a muon in a scattering medium and estimated MCS angle distribution using 2D Gaussian approximation. Red and blue represent high and low intensity of muon scattering angles, respectively [23].

**Point-of-Closest Approach**

To reconstruct three-dimensional images of target objects using muon scattering tomography, it is necessary to locate a muon scattering point within a volume of interest. Although muon scattering trajectories can be estimated using Monte-Carlo simulations and Bayesian estimation [24], [25], it is not possible to reconstruct the actual muon scattering trajectory in the medium. To address this limitation, a Point-of-Closest Approach (PoCA) algorithm which is simple and fast, is frequently used to identify a single muon scattering position. In the PoCA algorithm, only a single muon scattering point is assigned using reconstructed incoming and outgoing muon trajectories. Both incoming and outgoing muon trajectories are typically reconstructed using two-fold muon trackers. A PoCA point is defined as the midpoint of the shortest line between straight incoming and outgoing trajectories as shown in Fig. 2.

**RESULTS**

**Momentum Integrated Muon Scattering Tomography**

Tomographic images of target objects are reconstructed based on muon scattering angle values and PoCA algorithm in MST. In this work, we present a new approach which mathematically couples muon scattering angle and momentum in a single value. For each muon event, a PoCA point, scattering angle, and momentum are recorded:

$$N_i = [P_x\ P_y\ P_z\ \theta\ p] \quad i = 1, \ldots, N_\mu, \qquad (3)$$

where $N_\mu$ is the total number of muon events, $P_{x,y,z}$ is the Cartesian coordinate of PoCA point, $\theta$ is the muon scattering angle, and $p$ is the muon momentum. In MMST, Eq (3) is updated to yield a new quantity, $M$, which is computed by a correlation:

$$M(p,\theta) \equiv \log_{10}\left(\frac{\theta\ [\text{rad}]}{p^k\ [\text{GeV/c}]}\right). \qquad (4)$$

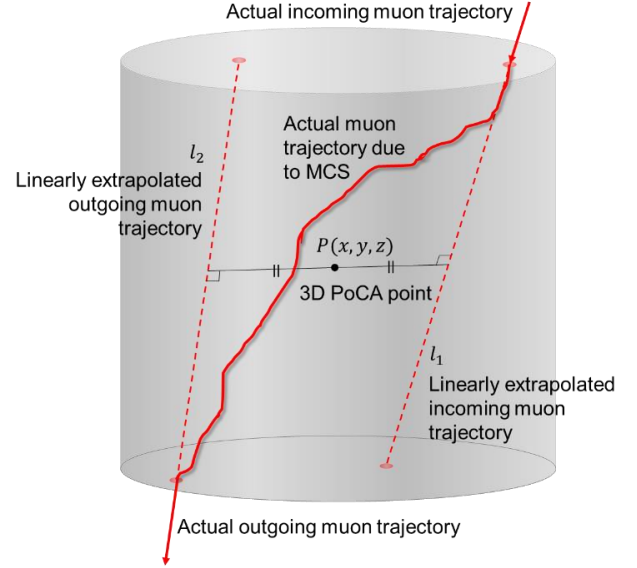

Fig. 2 A single muon scattering point estimation using a PoCA algorithm. Two extrapolated straight lines, incoming and outgoing muon trajectories, are reconstructed using two measured points in upper and lower muon trackers, respectively. The midpoint of the shortest perpendicular line is a PoCA point in 3D.

Although $k$ is determined by various factors, it can be estimated as a constant, $k = -2.4$ because its variance does not affect overall imaging quality. Details of derivation of Eq (4) and k value can be found in [23]. When M-value is computed using Eq (4), Eq (3) is updated by:

$$M_i = [P_x\ P_y\ P_z\ M] \quad i = 1, \ldots, N_\mu. \qquad (5)$$

**GEANT4 Simulations**

In this work, the Monte-Carlo particle transport simulation tool, GEANT4 (GEometry ANd Tracking) which was developed at CERN, is used to simulate muon transport and energy loss in materials. Using GEANT4, we built SNF in a dry cask canister model that contains up to 24 PWR fuel assemblies (FA). Each FA includes 15 × 15 $UO_2$ fuel rods. Diameter and length of a fuel rod are 10.7 and 3658 mm, respectively. A dimension of one FA is 2145 × 2145 × 3658 $mm^3$. FAs are surrounded by concrete shielding that has inner and outer radii of 863.5 and 1685 mm. In simulations, one of the middle FAs is excluded out of 24 PWR FAs as shown in Fig. 3.

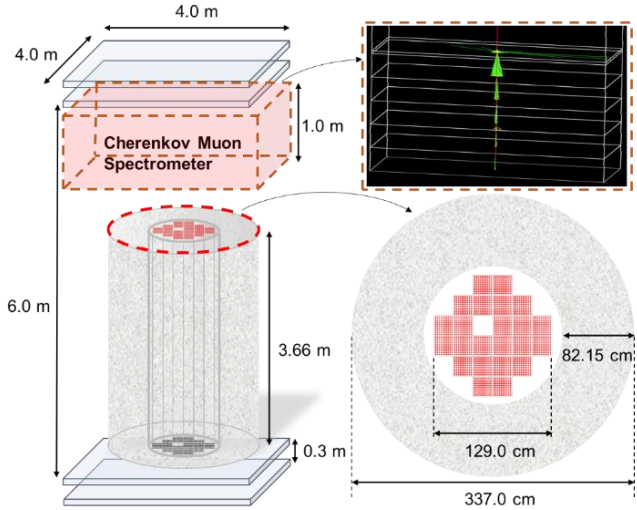

Fig. 3 Overview of momentum integrated muon scattering tomography (MMST) system using a Cherenkov muon spectrometer for SNF dry cask imaging (one FA is missing).

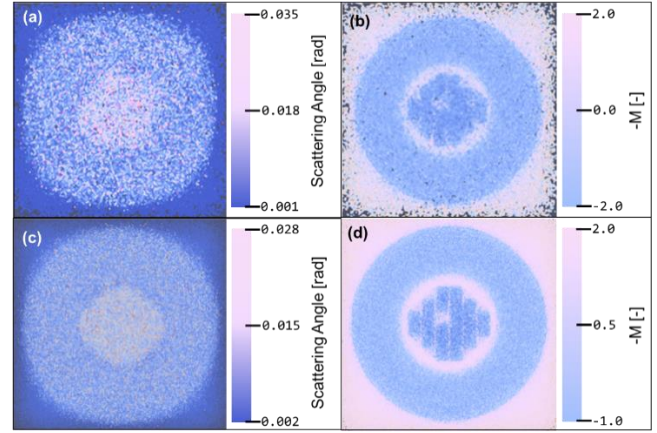

Fig. 4. Image reconstruction of SNF dry cask when one FA is missing using MST (left column) and MMST (right column) when $N_\mu = 10^5$ (upper row) and $10^6$ (lower row) muons.

In addition to SNFs in a dry cask, we also modeled muon trackers and a Cherenkov muon spectrometer to measure the scattering angle and momentum of muons in GEANT4 simulations. Muon trackers are made of $4 \times 4$ m$^2$ plastic scintillators. Distance between upper and lower muon trackers is 6.0 m. The Cherenkov muon spectrometer is placed below the upper tracker to measure muon momentum. SNF dry cask is located between muon spectrometer and trackers. Overall length and diameter of SNF cask are 3.66 m and 3.37 m, respectively. To simulate realistic energy spectrum and zenith angle distribution of cosmic ray muon, we incorporated the open-source Monte Carlo muon generator in GEANT4. Overview of the Cherenkov muon spectrometer and muon trackers for SNF dry cask imaging is shown in Fig. 3. In simulations, we generated $10^5$ and $10^6$ muons for monitoring SNF dry cask that represent 5–10 minutes and 1–2 hours of scanning time in the field condition.

**Monitoring Spent Nuclear Fuel in a Dry Cask**

The results of image reconstruction of SNF dry cask with one missing FA using MST and MMST when $N_\mu = 10^5$ and $10^6$ are shown in Fig. 4. It is noted that MST and MMST use a different color scale for imaging, scattering angle and M-value, respectively. Although finding the missing FA is challenging using MST (Fig. 4-(a) and (c)) due to low image resolution, we can locate it using MMST (Fig. 4-(b) and (d)). In addition to visual inspection, we also performed a systematical investigation to find a missing FA in the SNF dry cask. The results of systematical analyses to monitor SNF in a dry cask and find a central missing FA using MST and MMST when the numbers of muons are $10^5$ and $10^6$ are shown in Fig. 5. Various color plots are presented and they represent five (green), six (blue), and two FAs (red) at 215 mm, -215 mm, and -645 mm from the center of SNF assemblies, respectively. We enable to locate a missing FA using systematical analysis both using MST and MMST because of a noticeable drop in green plots in which a FA is absent in SNF assemblies in all cases. Although the uncertainty level (fluctuation in plots) of MST is improved by increasing the number of muons, it is still much higher than that of MMST.

**CONCLUSION**

In this paper, we present a new muon tomographic technique which allows momentum integration in muon scattering tomography. In MMST, a muon scattering angle value is substituted by a new quantity, $M$, that mathematically couples muon scattering angle and momentum in a single value. To demonstrate the feasibility of our proposed MMST, we use GEANT4 to simulate cosmic ray muon interactions within SNF in a dry cask. Specifically, we postulate a scenario that one of central FAs out of 24 SNF assemblies is missing. The results show that the location of the missing FA can be found using MMST in all cases whereas it is very challenging using MST even with 10 times more muons. Although the results are somewhat ideal, e.g., momentum and spatial resolution are $\sigma_p = 0.1$ GeV/c and $\sigma_x = 0.0$ mm, MMST appears to have the potential to improve MST by at least a factor of 10 in terms of investigating and monitoring the integrity of stored nuclear fuel in the cask.

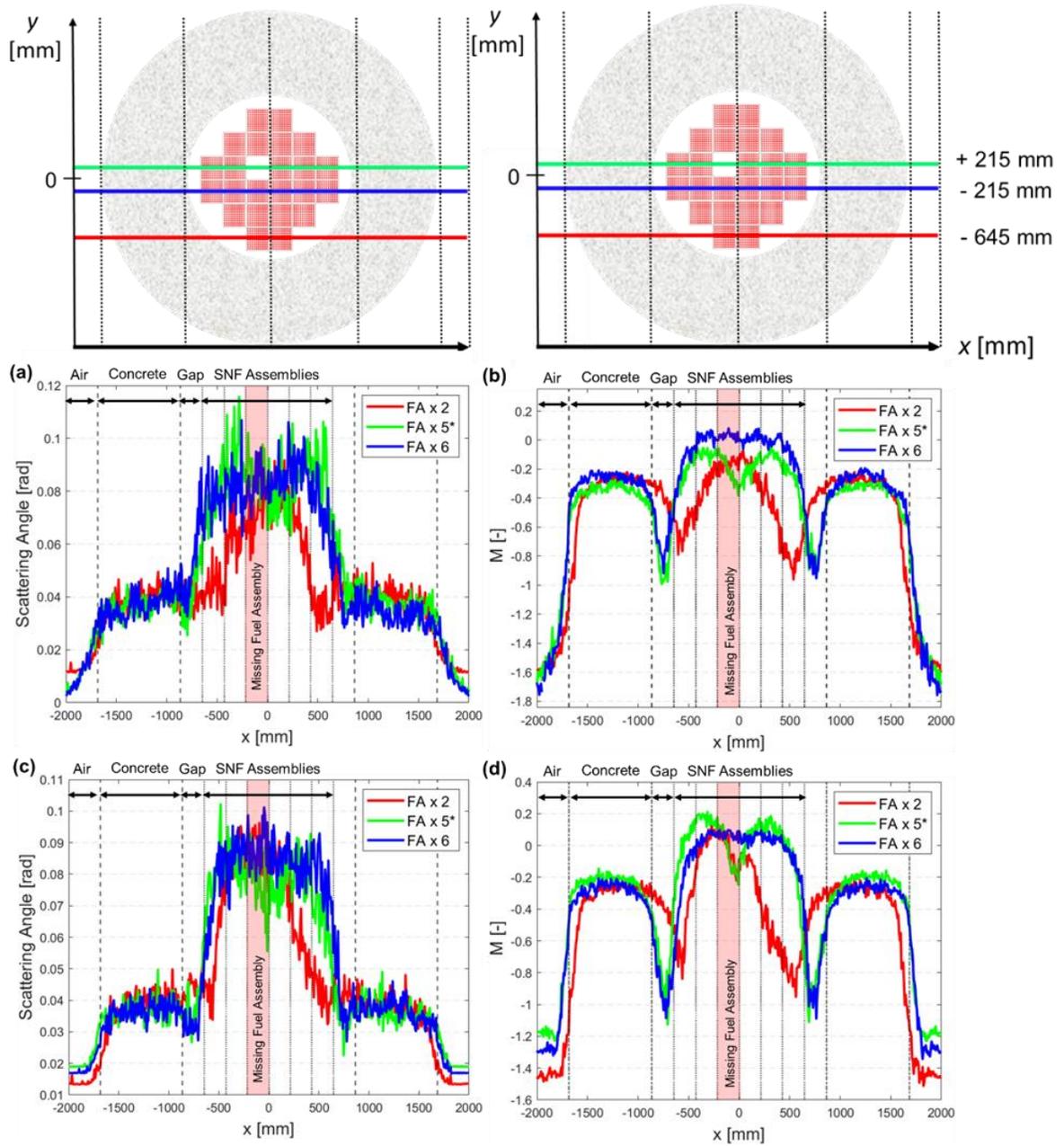

Fig. 5 Systematical analysis to monitor SNF in a dry cask to find a central missing FA (top) using MST (left) and MMST (right) when the numbers of muons are $10^5$ (middle) and $10^6$ (bottom). Various color plots are presented that represent 215 mm (green), -215 mm (blue), and -645 mm (red) from the center of SNF assemblies.

**ACKNOWLEDGMENT**

This research is being performed using funding from the Purdue Research Foundation.